\documentclass[10pt]{IEEEtran}
\usepackage[LGR,T1]{fontenc}
\usepackage[latin9]{inputenc}
\usepackage{amsmath}
\usepackage{amsthm}
\usepackage{amssymb}
\usepackage{graphicx}

\makeatletter
\theoremstyle{plain}
\newtheorem{thm}{\protect\theoremname}
\theoremstyle{plain}
\newtheorem{lem}[]{\protect\lemmaname}
\theoremstyle{remark}
\newtheorem{rem}[]{\protect\remarkname}
\theoremstyle{plain}
\newtheorem{cor}[]{\protect\corollaryname}

\usepackage{color}

\usepackage{cite}

\usepackage[mathscr]{eucal}
\usepackage{fixltx2e}
\usepackage{array}
\usepackage{verbatim}
\usepackage{bm}
\usepackage{algorithmic}
\usepackage{algorithm}
\usepackage{verbatim}
\usepackage{textcomp}
\usepackage{mathrsfs}
\usepackage{epstopdf}

\newcommand{\openone}{\leavevmode\hbox{\small1\normalsize\kern-.33em1}}

\catcode`~=11 \def\UrlSpecials{\do\~{\kern -.15em\lower .7ex\hbox{~}\kern .04em}} \catcode`~=13

\allowdisplaybreaks[4]

\newcommand{\calA}{\mathcal{A}}

\DeclareMathAlphabet{\mathbsf}{OT1}{cmss}{bx}{n}
\DeclareMathAlphabet{\mathssf}{OT1}{cmss}{m}{sl}

\DeclareSymbolFont{bsfletters}{OT1}{cmss}{bx}{n}
\DeclareSymbolFont{ssfletters}{OT1}{cmss}{m}{n}
\DeclareMathSymbol{\bsfGamma}{0}{bsfletters}{'000}
\DeclareMathSymbol{\ssfGamma}{0}{ssfletters}{'000}
\DeclareMathSymbol{\bsfDelta}{0}{bsfletters}{'001}
\DeclareMathSymbol{\ssfDelta}{0}{ssfletters}{'001}
\DeclareMathSymbol{\bsfTheta}{0}{bsfletters}{'002}
\DeclareMathSymbol{\ssfTheta}{0}{ssfletters}{'002}
\DeclareMathSymbol{\bsfLambda}{0}{bsfletters}{'003}
\DeclareMathSymbol{\ssfLambda}{0}{ssfletters}{'003}
\DeclareMathSymbol{\bsfXi}{0}{bsfletters}{'004}
\DeclareMathSymbol{\ssfXi}{0}{ssfletters}{'004}
\DeclareMathSymbol{\bsfPi}{0}{bsfletters}{'005}
\DeclareMathSymbol{\ssfPi}{0}{ssfletters}{'005}
\DeclareMathSymbol{\bsfSigma}{0}{bsfletters}{'006}
\DeclareMathSymbol{\ssfSigma}{0}{ssfletters}{'006}
\DeclareMathSymbol{\bsfUpsilon}{0}{bsfletters}{'007}
\DeclareMathSymbol{\ssfUpsilon}{0}{ssfletters}{'007}
\DeclareMathSymbol{\bsfPhi}{0}{bsfletters}{'010}
\DeclareMathSymbol{\ssfPhi}{0}{ssfletters}{'010}
\DeclareMathSymbol{\bsfPsi}{0}{bsfletters}{'011}
\DeclareMathSymbol{\ssfPsi}{0}{ssfletters}{'011}
\DeclareMathSymbol{\bsfOmega}{0}{bsfletters}{'012}
\DeclareMathSymbol{\ssfOmega}{0}{ssfletters}{'012}

\DeclareMathOperator{\supp}{supp}

\usepackage{graphicx}

\def\e{{\rm e}}

\allowdisplaybreaks[1]
\providecommand{\corollaryname}{Corollary}

\providecommand{\lemmaname}{Lemma}
\providecommand{\remarkname}{Remark}
\providecommand{\theoremname}{Theorem}

\providecommand{\lemmaname}{Lemma}
\providecommand{\theoremname}{Theorem}

\makeatother

\providecommand{\corollaryname}{Corollary}
\providecommand{\lemmaname}{Lemma}
\providecommand{\remarkname}{Remark}
\providecommand{\theoremname}{Theorem}

\begin{document}
\title{Corrections to ``Wyner's Common Information under R\'enyi Divergence
Measures''}
\author{Lei Yu and Vincent Y. F. Tan, \IEEEmembership{Senior Member,~IEEE}
\thanks{ L.~Yu is with the Department of Electrical Engineering and Computer Sciences, University of California, Berkeley, CA 94720, USA (e-mail: leiyu@berkeley.edu). V.~Y.~F.~Tan is with the Department of Electrical and Computer Engineering and the Department of Mathematics, National University of Singapore, Singapore 119076 (e-mail: vtan@nus.edu.sg).} \thanks{ Communicated by M. Raginsky, Associate Editor for Probability and Statistics. } \thanks{Copyright (c) 2019 IEEE. Personal use of this material is permitted. However, permission to use this material for any other purposes must be obtained from the IEEE by sending a request to pubs-permissions@ieee.org.} }
\maketitle
\begin{abstract}
In this correspondence, we correct an erroneous result on the achievability
part of the R\'enyi common information with order $1+s\in(1,2]$ in
\cite{yu2018wyner}.  The new achievability result (upper bound)
of the R\'enyi common information no longer coincides with Wyner's common
information. We also provide a new converse result (lower bound)
in this correspondence for the R\'enyi common information with order
$1+s\in(1,\infty]$. Numerical results show that for doubly symmetric
binary sources, the new upper and lower bounds coincide for the order
$1+s\in(1,2]$ and they are both strictly larger than Wyner's common
information for this case.
\end{abstract}

\section{Introduction}

In the paper \cite{yu2018wyner}, we defined a new notion, the R\'enyi
common information, which is a generalization of Wyner's common information.
This generalization involves using the unnormalized and normalized
R\'enyi divergences, instead of the relative entropy, to measure the
level of approximation between the induced and target distributions.
For a given target distribution $\pi_{XY}$, the minimum rate needed
to ensure that the unnormalized R\'enyi divergence $D_{1+s}(P_{X^{n}Y^{n}}\|\pi_{X^{n}Y^{n}})$
(resp. normalized R\'enyi divergence $\frac{1}{n}D_{1+s}(P_{X^{n}Y^{n}}\|\pi_{X^{n}Y^{n}})$)
vanishes asymptotically is defined as the R\'enyi common information,
and denoted as $T_{1+s}(\pi_{XY})$ (resp. $\widetilde{T}_{1+s}(\pi_{XY})$).
Here $\pi_{X^{n}Y^{n}}:=\pi_{XY}^{n}$. The case of $s=0$ corresponds
to Wyner's common information, which is equal to\footnote{In \cite{yu2018wyner}, $C_{\mathsf{Wyner}}(\pi_{XY})$ was denoted
as $C_{\mathsf{Wyner}}(X;Y)$.} $C_{\mathsf{Wyner}}(\pi_{XY})$ defined in \cite[Eqn. (1)]{yu2018wyner}.
In \cite{yu2018wyner}, we focused on the cases that $\pi_{XY}$ has
a finite alphabet and the R\'enyi parameter $1+s\in[0,2]$. In Theorem
1 of \cite{yu2018wyner}, we claimed that for these cases, the R\'enyi
common information was equal to Wyner's common information. However,
in fact, this is incorrect. There is an error in the achievability
proof part of Theorem 1 for $s\in(0,1]$. Obviously, for $s\in(0,1]$,
by definition, $T_{1+s}(\pi_{XY})$ and $\widetilde{T}_{1+s}(\pi_{XY})$
are lower bounded by $C_{\mathsf{Wyner}}(\pi_{XY})$. But the proof
for that they are upper bounded by $C_{\mathsf{Wyner}}(\pi_{XY})$
(the achievability part) for the case $s\in(0,1]$ is incorrect. Specifically,
in the proof given in Appendix A of \cite{yu2018wyner}, equation
(79) is incorrect, since for a tuple $(w^{n},x^{n},y^{n})$, the conditions
that $(w^{n},x^{n})$ has joint type $T_{W}V_{X|W}$ and $(w^{n},y^{n})$
has joint type $T_{W}V_{Y|W}$ do not necessarily imply that $(x^{n},y^{n})$
has joint type $\sum_{w}T_{W}(w)V_{X|W}(\cdot|w)V_{Y|W}(\cdot|w)$.
In fact, the type of $(x^{n},y^{n})$ can be any element of the set
\begin{align}
 & \Bigl\{\sum_{w}T_{W}(w)V_{XY|W}'(\cdot|w):\:T_{W}V_{XY|W}'\textrm{ is a type s.t. }\nonumber \\
 & \qquad V_{X|W}'=V_{X|W},V_{Y|W}'=V_{Y|W}\Bigr\}.
\end{align}
 In this document, we correct the erroneous statement in Theorem 1
of \cite{yu2018wyner} and provide a corresponding proof of the new,
albeit weaker, claim.

Denote the \emph{coupling sets} of $(P_{X},P_{Y})$ and $(P_{X|W},P_{Y|W})$
respectively as 
\begin{align}
C(P_{X},P_{Y}) & :=\bigl\{ Q_{XY}\in\mathcal{P}(\mathcal{X}\times\mathcal{Y}):\nonumber \\
 & \qquad Q_{X}=P_{X},Q_{Y}=P_{Y}\bigr\},\\
C(P_{X|W},P_{Y|W}) & :=\bigl\{ Q_{XY|W}\in\mathcal{P}(\mathcal{X}\times\mathcal{Y}|\mathcal{W}):\nonumber \\
 & \qquad Q_{X|W}=P_{X|W},Q_{Y|W}=P_{Y|W}\bigr\}.\label{eq:-33-2}
\end{align}
Define the \emph{maximal $s$-mixed Shannon-cross entropy} with respect
to $\pi_{XY}$ over couplings $C(P_{X},P_{Y})$ as\footnote{Throughout this paper, we use $H(Q_{X})$ or $H_{Q}(X)$ to denote
the entropy of $X\sim Q_{X}$. When the distribution is denoted by
$P_{X}$, we omit the subscript, i.e., $H(X):=H_{P}(X)$. This notation
convention also applies to the conditional entropy and mutual information.} 
\begin{align}
 & \mathcal{H}_{s}(P_{X},P_{Y}\|\pi_{XY})\nonumber \\
 & :=\max_{Q_{XY}\in C(P_{X},P_{Y})}\sum_{x,y}Q_{XY}(x,y)\log\frac{1}{\pi\left(x,y\right)}+\frac{1}{s}H(Q_{XY}).\label{eq:maximalcrossentropy}
\end{align}
For $s\in(0,\infty)$, define
\begin{align}
\Gamma_{1+s}^{\mathrm{UB}}(\pi_{XY}) & :=\min_{P_{W}P_{X|W}P_{Y|W}:P_{XY}=\pi_{XY}}-\frac{1+s}{s}H(XY|W)\nonumber \\
 & \qquad+\sum_{w}P(w)\mathcal{H}_{s}(P_{X|W=w},P_{Y|W=w}\|\pi_{XY})\label{eq:-35-4}
\end{align}
and 
\begin{align}
\Gamma_{1+s}^{\mathrm{LB}}(\pi_{XY}) & :=\inf_{P_{W}P_{X|W}P_{Y|W}:P_{XY}=\pi_{XY}}-\frac{1+s}{s}H(XY|W)\nonumber \\
 & \qquad+\inf_{Q_{WW'}\in C(P_{W},P_{W})}\sum_{w,w'}Q(w,w')\nonumber \\
 & \qquad\times\mathcal{H}_{s}(P_{X|W=w},P_{Y|W=w'}\|\pi_{XY}).\label{eq:-35-4-1}
\end{align}
Define $\Gamma_{1}^{\mathrm{UB}}(\pi_{XY}),\Gamma_{1}^{\mathrm{LB}}(\pi_{XY}),\Gamma_{\infty}^{\mathrm{UB}}(\pi_{XY}),$
and $\Gamma_{\infty}^{\mathrm{LB}}(\pi_{XY})$ as the continuous extensions
of $\Gamma_{1+s}^{\mathrm{UB}}(\pi_{XY})$ and $\Gamma_{1+s}^{\mathrm{LB}}(\pi_{XY})$
as $s$ tends to $0$ or $\infty$.

We introduce a condition on the distribution $\pi_{XY}$, which will
be used to characterize the necessary and sufficient condition for
$\Gamma_{1+s}^{\mathrm{UB}}(\pi_{XY})=C_{\mathrm{Wyner}}(\pi_{XY})$.

Condition $(*)$: There exists some optimal distribution $P_{W}P_{X|W}P_{Y|W}$
attaining $C_{\mathrm{Wyner}}(\pi_{XY})$ such that $\pi_{XY}$ is
product on $\supp\left(P_{X|W=w}\right)\times\supp\left(P_{Y|W=w}\right)$
for each $w\in\supp\left(P_{W}\right)$, i.e., $\pi_{XY}\left(\cdot|\supp\left(P_{X|W=w}\right)\times\supp\left(P_{Y|W=w}\right)\right)$
is a product distribution for each $w\in\supp\left(P_{W}\right)$.

Now we provide some useful properties of $\Gamma_{1+s}^{\mathrm{UB}}(\pi_{XY})$
and $\Gamma_{1+s}^{\mathrm{LB}}(\pi_{XY})$.
\begin{lem}
\label{lem:property} 1) In \eqref{eq:-35-4}, it suffices to restrict
the alphabet size of $W$ such that $|\mathcal{W}|\le|\mathcal{X}||\mathcal{Y}|$.
\\
2) $\Gamma_{1+s}^{\mathrm{UB}}(\pi_{XY})$ and $\Gamma_{1+s}^{\mathrm{LB}}(\pi_{XY})$
are non-decreasing in $s\in(0,\infty)$.\\
3) The following limiting cases hold. 
\begin{align}
\Gamma_{1}^{\mathrm{LB}}(\pi_{XY}) & \leq\Gamma_{1}^{\mathrm{UB}}(\pi_{XY})=C_{\mathsf{Wyner}}(X;Y),\label{eq:-42}\\
\Gamma_{\infty}^{\mathrm{UB}}(\pi_{XY}) & =\min_{\substack{P_{W}P_{X|W}P_{Y|W}:\\
P_{XY}=\pi_{XY}
}
}-H(XY|W)+\sum_{w}P(w)\nonumber \\
 & \quad\times\max_{\substack{Q_{XY}\in\\
C(P_{X|W=w},P_{Y|W=w})
}
}\sum_{x,y}Q(x,y)\log\frac{1}{\pi\left(x,y\right)},\label{eq:-36-5}\\
\Gamma_{\infty}^{\mathrm{LB}}(\pi_{XY}) & =\inf_{\substack{P_{W}P_{X|W}P_{Y|W}:\\
P_{XY}=\pi_{XY}
}
}-H(XY|W)\nonumber \\
 & \quad+\inf_{\substack{Q_{WW'}\in\\
C(P_{W},P_{W})
}
}\sum_{w,w'}Q(w,w')\nonumber \\
 & \quad\times\max_{\substack{Q_{XY}\in\\
C(P_{X|W=w},P_{Y|W=w'})
}
}\sum_{x,y}Q(x,y)\log\frac{1}{\pi\left(x,y\right)}.\label{eq:-41}
\end{align}
4) For $s\in(0,\infty]$, $\Gamma_{1+s}^{\mathrm{UB}}(\pi_{XY})=C_{\mathrm{Wyner}}(\pi_{XY})$
if and only if $\pi_{XY}$ satisfies the condition $(*)$.
\end{lem}
The proof of Lemma \ref{lem:property} is provided in Appendix \ref{sec:Proof-of-Lemma-property}.
Now we provide the promised correction of \cite[Theorem 1]{yu2018wyner}.
\begin{thm}[R\'enyi Common Informations]
\label{thm:RenyiCI} The unnormalized and normalized and R\'enyi common
informations satisfy 
\begin{align}
\widetilde{T}_{1+s}(\pi_{XY}) & =T_{1+s}(\pi_{XY})\\
 & =\begin{cases}
C_{\mathsf{Wyner}}(X;Y) & s\in(-1,0]\\
0 & s=-1
\end{cases},
\end{align}
\begin{align}
T_{1+s}(\pi_{XY}) & \geq\widetilde{T}_{1+s}(\pi_{XY})\nonumber \\
 & \geq\max\left\{ \Gamma_{1+s}^{\mathrm{LB}}(\pi_{XY}),C_{\mathsf{Wyner}}(\pi_{XY})\right\} ,\;s\in(0,\infty],\label{eq:-28}
\end{align}
and 
\begin{align}
 & \widetilde{T}_{1+s}(\pi_{XY})\leq T_{1+s}(\pi_{XY})\leq\Gamma_{1+s}^{\mathrm{UB}}(\pi_{XY}),\;s\in(0,1]\cup\{\infty\}.
\end{align}
Furthermore, for $s\in(-1,1]\cup\{\infty\}$, the optimal R\'enyi divergence
$D_{1+s}(P_{X^{n}Y^{n}}\|\pi_{X^{n}Y^{n}})$ in the definitions of
the R\'enyi common informations decays at least exponentially fast in
$n$ when $R>C_{\mathsf{Wyner}}(X;Y)$ for $s\in(-1,0]$ and $R>\Gamma_{1+s}^{\mathrm{UB}}(\pi_{XY})$
for $s\in(0,1]\cup\{\infty\}$.
\end{thm}
\begin{rem}
By Statement 4) of Lemma \ref{lem:property}, we know that for any
pseudo-product distribution $\pi_{XY}$, the (unnormalized and normalized)
R\'enyi common informations with $s\in(-1,\infty]$ are equal to Wyner's
common information, i.e., 
\begin{equation}
\widetilde{T}_{1+s}(\pi_{XY})=T_{1+s}(\pi_{XY})=C_{\mathrm{Wyner}}(\pi_{XY}),\forall s\in(-1,\infty].
\end{equation}
\end{rem}
The upper bound for the case $s\in(0,1]$ is proved in Section \ref{sec:Upper-Bound-for}.
The lower bound for the case $s\in(0,\infty]$ is proved in Section
\ref{sec:Lower-Bound-for}. The upper and lower bounds for the case
$s=\infty$ were derived by the present authors in \cite{yu2018on}.
Hence for the achievability part, here we only provide a proof for
$s\in(0,1]$. (The converse proof that we present here includes the
case $s=\infty$).

\begin{figure*}
\centering \includegraphics[width=0.6\textwidth]{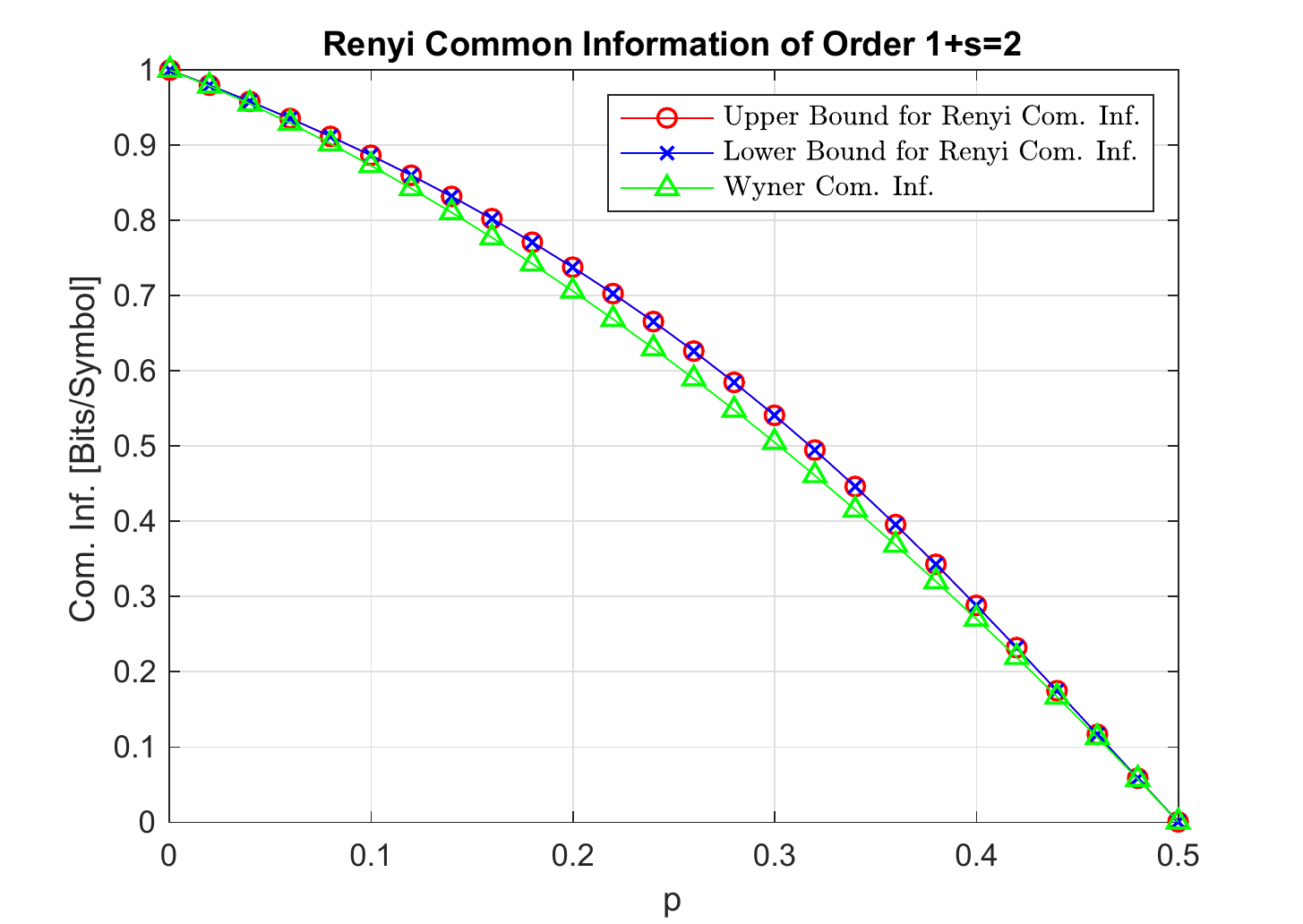}

\centering \includegraphics[width=0.6\textwidth]{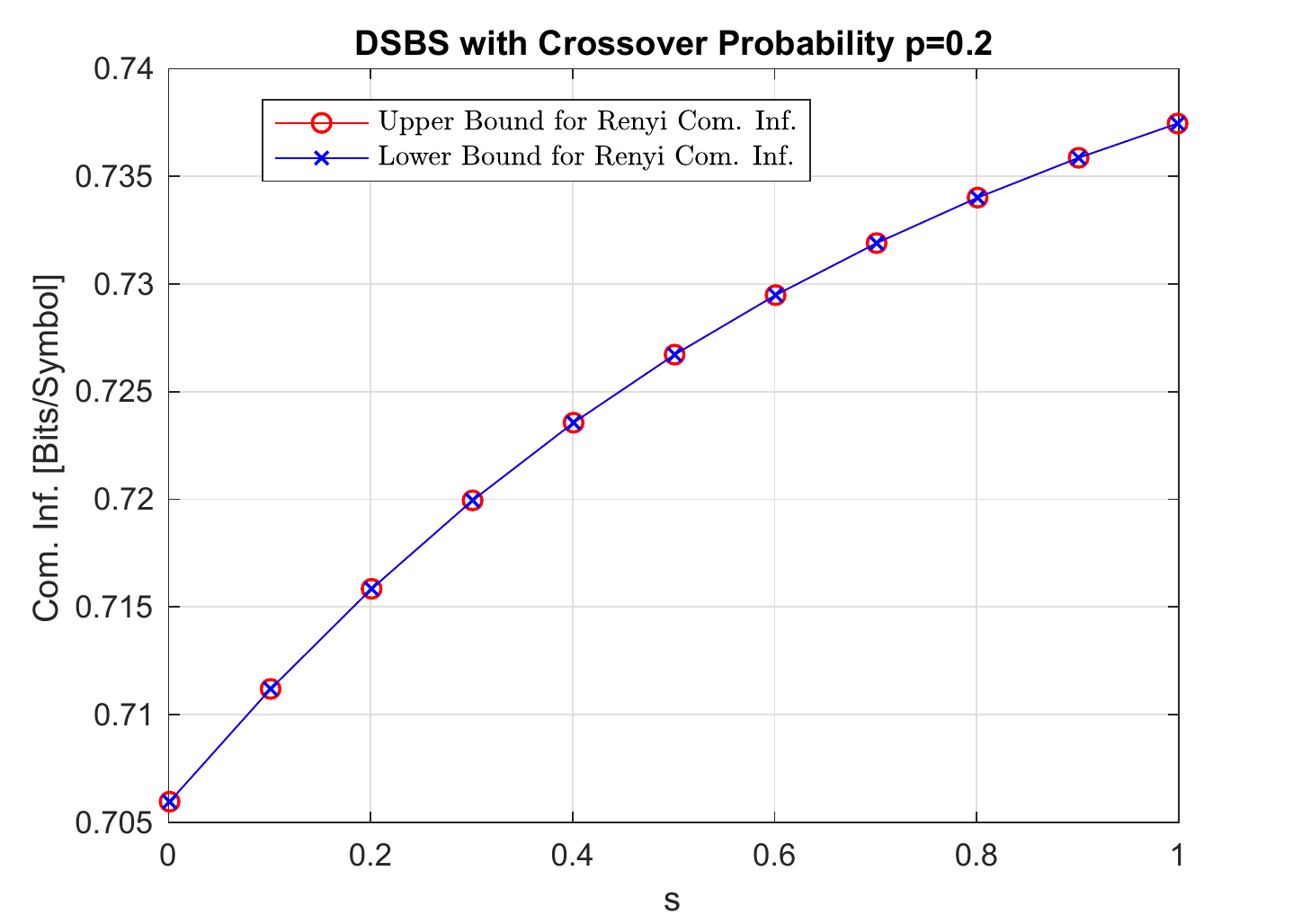}

\caption{\label{fig:Common-informations-for}Illustrations of the upper bound
in \eqref{eq:-26} and lower bound in \eqref{eq:-28} for the R\'enyi
common informations, as well as Wyner's common information in \eqref{eq:-25}
for the DSBS $(X,Y)$ such that $X\sim\mathrm{Bern}(\frac{1}{2})$
and $Y=X\oplus E$ with $E\sim\mathrm{Bern}(p)$ independent of $X$.
In the top figure, $s=1$; and in the bottom figure, $p=0.2$ (i.e.,
$\alpha_{0}=0.4$). For the lower bound, we gradually increase the
alphabet size of $W$ from $2$ to $10$. Numerical results show that
the resulting curve does not change when we increase the alphabet
size. That is, for the DSBS, it appears that restricting the alphabet
size of $W$ to $2$ suffices to attain the lower bound. }
\end{figure*}

To illustrate that the upper bound $\Gamma_{1+s}^{\mathrm{UB}}(\pi_{XY})$
and the lower bound $\Gamma_{1+s}^{\mathrm{LB}}(\pi_{XY})$ may coincide
for certain sources, we now consider a doubly symmetric binary source
(DSBS) $\left(X,Y\right)$ with joint distribution 
\begin{equation}
\pi_{XY}:=\left[\begin{array}{cc}
\alpha_{0} & \beta_{0}\\
\beta_{0} & \alpha_{0}
\end{array}\right]
\end{equation}
where $\alpha_{0}=\frac{1}{2}\left(a^{2}+(1-a)^{2}\right),\beta_{0}=a(1-a)$
with $a\in(0,\frac{1}{2})$. That is equivalent to the setting that
$W\sim\mathrm{Bern}(\frac{1}{2})$, $X=W\oplus A$, $Y=W\oplus B$,
$A\sim\mathrm{Bern}(a)$ and $B\sim\mathrm{Bern}(a)$ are independent.
Then by using Theorem \ref{thm:RenyiCI}, we can obtain the following
results.
\begin{cor}
\label{cor:For-a-DSBS}For a DSBS $\left(X,Y\right)$ with distribution
$\pi_{XY}$, we have that:\\
1) For $s\in(-1,0]$,
\begin{align}
 & \widetilde{T}_{1+s}(\pi_{XY})\nonumber \\
 & =T_{1+s}(\pi_{XY})\\
 & =-2H_{2}(a)-\left(a^{2}+(1-a)^{2}\right)\log\left[\frac{a^{2}+(1-a)^{2}}{2}\right]\nonumber \\
 & \qquad-2a(1-a)\log\left[a(1-a)\right],
\end{align}
where $H_{2}(a):=-a\log a-(1-a)\log(1-a)$ denotes the binary entropy
function.\\
2) For $s\in(0,1]$, 
\begin{align}
 & \widetilde{T}_{1+s}(\pi_{XY})\nonumber \\
 & \leq T_{1+s}(\pi_{XY})\\
 & \leq-\frac{1+s}{s}2H_{2}(a)+\frac{1}{s}\bigl\{-p^{*}\log p^{*}-2(a-p^{*})\log(a-p^{*})\nonumber \\
 & \qquad-(1+p^{*}-2a)\log(1+p^{*}-2a)\nonumber \\
 & \qquad-s(1+2p^{*}-2a)\log\alpha_{0}-s(2a-2p^{*})\log\beta_{0}\bigr\},\label{eq:-26}
\end{align}
where $p^{*}=\frac{\sqrt{k^{2}\left(\overline{a}-a\right)^{2}+4ka\overline{a}}-\left(k\left(\overline{a}-a\right)+2a\right)}{2\left(k-1\right)}$
and $k=\left(\frac{\alpha_{0}}{\beta_{0}}\right)^{2s}$.\\
3) For $s=\infty$,
\begin{align}
\widetilde{T}_{\infty}(\pi_{XY}) & =T_{\infty}(\pi_{XY})\\
 & =-2H_{2}(a)-(1-2a)\log\left[\frac{a^{2}+(1-a)^{2}}{2}\right]\nonumber \\
 & \qquad-2a\log\left[a(1-a)\right].\label{eq:-27}
\end{align}
\end{cor}
\begin{IEEEproof}
For the DSBS, Wyner \cite{Wyner} showed that 
\begin{align}
 & C_{\mathrm{Wyner}}(\pi_{XY})\nonumber \\
 & =-2H_{2}(a)-\left(a^{2}+(1-a)^{2}\right)\log\left[\frac{1}{2}\left(a^{2}+(1-a)^{2}\right)\right]\nonumber \\
 & \qquad-2a(1-a)\log\left[a(1-a)\right].\label{eq:-25}
\end{align}
Combining this with Theorem \ref{thm:RenyiCI}, we obtain Statement
1).

For Statement 2), we set $W\sim\mathrm{Bern}(\frac{1}{2})$, $X=W\oplus A$,
and $Y=W\oplus B$, where $A,B\sim\mathrm{Bern}(a)$ with $a\in(0,\frac{1}{2})$
are independent. For this setting, 
\begin{align}
 & C(P_{X|W=w},P_{Y|W=w})\nonumber \\
 & =\left\{ \left[\begin{array}{cc}
p & a-p\\
a-p & 1+p-2a
\end{array}\right]:0\le p\le a\right\} .
\end{align}
Therefore,
\begin{align}
 & s\mathcal{H}_{s}(P_{X|W=w},P_{Y|W=w}\|\pi_{XY})\nonumber \\
 & =\max_{\substack{Q_{XY}\in\\
C(P_{X|W=w},P_{Y|W=w})
}
}\sum_{x,y}Q(x,y)\log\frac{1}{\pi\left(x,y\right)^{s}Q(x,y)}\label{eq:-54}\\
 & =-p^{*}\log p^{*}-2(a-p^{*})\log(a-p^{*})\nonumber \\
 & \qquad-(1+p^{*}-2a)\log(1+p^{*}-2a)\nonumber \\
 & \qquad-s(1+2p^{*}-2a)\log\alpha_{0}-s(2a-2p^{*})\log\beta_{0}\Bigr\},
\end{align}
where the optimal $Q_{XY}$ in \eqref{eq:-54} is 
\begin{equation}
\left[\begin{array}{cc}
p^{*} & a-p^{*}\\
a-p^{*} & 1+p^{*}-2a
\end{array}\right].
\end{equation}
Hence $\Gamma_{1+s}^{\mathrm{UB}}(\pi_{XY})$ is upper bounded by
the expression in \eqref{eq:-26}. Combining this with Theorem \ref{thm:RenyiCI},
$T_{1+s}(\pi_{XY})$ is also upper bounded by the expression in \eqref{eq:-26}.

Statement 3) was proven in \cite[Theorem 3]{yu2018on}. 
\end{IEEEproof}
The upper bound and lower bound for the R\'enyi common informations,
as well as Wyner's common information for the DSBS are illustrated
in Fig.~\ref{fig:Common-informations-for}.

It is easy to verify that the expression in \eqref{eq:-27} (and also
the upper bound in Corollary \ref{cor:For-a-DSBS} for $s>0$) is
strictly larger than the expression in \eqref{eq:-25}. Hence, for
the DSBS, the R\'enyi common information of order $\infty$ is strictly
larger than Wyner's common information; also see \cite[Corollary 1]{yu2018on}.
Furthermore, when we evaluate the lower bound $\Gamma_{1+s}^{\mathrm{LB}}(\pi_{XY})$
for the DSBS, the numerical results (in Fig.~\ref{fig:Common-informations-for})
show that the lower bound coincides with the upper bound in Corollary
\ref{cor:For-a-DSBS}. Hence it is natural to conjecture the upper
bound in Corollary \ref{cor:For-a-DSBS} for the DSBS is tight. To
show this, one may follow the proof idea used in \cite[Theorem 3]{yu2018on}
for the case $s=\infty$. However, for $s\in(0,\infty)$, the proof
is not straightforward and we leave this to future work. 

\section{\label{sec:Upper-Bound-for}Upper Bound for Case $s\in(0,1]$}
\begin{IEEEproof}
Here we only prove that $T_{1+s}(\pi_{XY})\leq\Gamma_{1+s}^{\mathrm{UB}}(\pi_{XY})$
for $s\in(0,1]$. Other parts have been proven in the original paper
\cite{yu2018wyner}.

We continue from \cite[Eqn.~(74)]{yu2018wyner}. Analogously to the
definition of ${\cal A}$ in (74), here we define 
\begin{align}
 & \calA_{\epsilon}':=\Bigl\{ P_{XY|W}\in\mathcal{P}(\mathcal{X}\times\mathcal{Y}|\mathcal{W}):\nonumber \\
 & \forall(w,x),\left|\left[Q_{W}P_{X|W}\right](w,x)-Q_{WX}(w,x)\right|\leq\epsilon Q_{WX}(w,x),\nonumber \\
 & \forall(w,y),\left|\left[Q_{W}P_{Y|W}\right](w,y)-Q_{WY}(w,y)\right|\leq\epsilon Q_{WY}(w,y)\Bigr\}.
\end{align}
Note that in \cite[Eqn.~(74)]{yu2018wyner}, we omit the dependence
of ${\cal A}$ on $\epsilon$. It is also worth noting that the set
${\cal A}$ defined in \cite[Eqn.~(74)]{yu2018wyner} can be written
as
\begin{align}
{\cal A} & =\Bigl\{ T_{w^{n}x^{n}y^{n}}:w^{n}\in\mathcal{T}_{\epsilon'}^{n}\left(Q_{W}\right),\nonumber \\
 & \qquad x^{n}\in\mathcal{T}_{\epsilon}^{n}\left(Q_{WX}|w^{n}\right),y^{n}\in\mathcal{T}_{\epsilon}^{n}\left(Q_{WY}|w^{n}\right)\Bigr\},
\end{align}
where $T_{w^{n}x^{n}y^{n}}$ denotes the joint type of $(w^{n},x^{n},y^{n})$.

Let 
\begin{align}
\delta_{0,n} & :=1-Q_{W}^{n}\left(\mathcal{T}_{\epsilon'}^{n}\left(Q_{W}\right)\right),\\
\delta_{1,n} & :=1-\min_{w^{n}\in\mathcal{T}_{\epsilon'}^{n}}Q_{X|W}^{n}\bigl(\mathcal{T}_{\epsilon}^{n}(Q_{WX}|w^{n})|w^{n}\bigr),\\
\delta_{2,n} & :=1-\min_{w^{n}\in\mathcal{T}_{\epsilon'}^{n}}Q_{Y|W}^{n}\bigl(\mathcal{T}_{\epsilon}^{n}(Q_{WY}|w^{n})|w^{n}\bigr).
\end{align}
By \cite[Lemma 4]{yu2018wyner}, $\delta_{0,n},\delta_{1,n},\delta_{2,n}\to0$
as $n\to\infty$. Let 
\begin{equation}
\delta_{012,n}:=\frac{1}{ns}\log\left[(1-\delta_{0,n})(1-\delta_{1,n})^{1+s}(1-\delta_{2,n})^{1+s}\right].
\end{equation}
Hence $\delta_{012,n}\to0$ as $n\to\infty$. 

Consider that \eqref{eq:-24-1}-\eqref{eq:-38} (given on the page
\pageref{eq:-24-1}), where \eqref{eq:-32} follows from the definition
of typical sets and the definitions of ${\cal A},\delta_{0,n},\delta_{1,n},\delta_{2,n}$,
\eqref{eq:-8} follows since $\left|\mathcal{T}_{T_{WXY}}\right|\le\e^{nH(T_{WXY})}$
(see \cite[Lemma 2.3]{Csiszar}), in \eqref{eq:-9-1}, $\delta_{n}:=\frac{1}{n}\log|{\cal A}|\to0$
as $n\rightarrow\infty$ since the number of types is polynomial in
$n$ (see \cite[Lemma 2.2]{Csiszar}), and \eqref{eq:-38} follows
since $D(T_{W}\|Q_{W})\ge0$ and $\left\{ P_{XY|W}:\ \exists P_{W}\;\textrm{s.t.}\;P_{XY|W}P_{W}\in{\cal A}\right\} \subseteq\calA_{\epsilon}'.$

\begin{figure*}[t]
\begin{align}
 & \frac{1}{n}D_{1+s}\left(P_{W^{n}X^{n}Y^{n}}\|P_{W^{n}}\pi_{X^{n}Y^{n}}\right)\nonumber \\
 & =\frac{1}{ns}\log\sum_{w^{n},x^{n},y^{n}}P\left(w^{n}\right)\left(P\left(x^{n}|w^{n}\right)P\left(y^{n}|w^{n}\right)\right)^{1+s}\pi^{-s}(x^{n},y^{n})\label{eq:-24-1}\\
 & =\frac{1}{ns}\log\sum_{T_{WXY}}\sum_{\substack{(w^{n},x^{n},y^{n})\in\mathcal{T}_{T_{WXY}}}
}\frac{Q_{W}^{n}\left(w^{n}\right)1\left\{ w^{n}\in\mathcal{T}_{\epsilon'}^{n}\left(Q_{W}\right)\right\} }{Q_{W}^{n}\left(\mathcal{T}_{\epsilon'}^{n}\left(Q_{W}\right)\right)}\nonumber \\*
 & \qquad\times\left(\frac{Q_{X|W}^{n}\left(x^{n}|w^{n}\right)1\left\{ x^{n}\in\mathcal{T}_{\epsilon}^{n}\left(Q_{WX}|w^{n}\right)\right\} }{Q_{X|W}^{n}\left(\mathcal{T}_{\epsilon}^{n}\left(Q_{WX}|w^{n}\right)|w^{n}\right)}\frac{Q_{Y|W}^{n}\left(y^{n}|w^{n}\right)1\left\{ y^{n}\in\mathcal{T}_{\epsilon}^{n}\left(Q_{WY}|w^{n}\right)\right\} }{Q_{Y|W}^{n}\left(\mathcal{T}_{\epsilon}^{n}\left(Q_{WY}|w^{n}\right)|w^{n}\right)}\right)^{1+s}\pi_{X^{n}Y^{n}}^{-s}(x^{n},y^{n})\\
 & \leq\frac{1}{ns}\log\sum_{T_{WXY}\in{\cal A}}\sum_{\substack{(w^{n},x^{n},y^{n})\in\mathcal{T}_{T_{WXY}}}
}\frac{\e^{n\sum_{w,x}T\left(w\right)\log Q\left(w\right)}}{1-\delta_{0,n}}\nonumber \\
 & \qquad\times\left(\frac{\e^{n\sum_{w,x}T\left(w,x\right)\log Q\left(x|w\right)}}{1-\delta_{1,n}}\frac{\e^{n\sum_{w,y}T\left(w,y\right)\log Q\left(y|w\right)}}{1-\delta_{2,n}}\right)^{1+s}\e^{-ns\sum_{x,y}T\left(x,y\right)\log\pi\left(x,y\right)}\label{eq:-32}\\
 & \leq-\delta_{012,n}+\frac{1}{ns}\log\sum_{T_{WXY}\in{\cal A}}\e^{nH(T_{WXY})+n\sum_{w,x}T\left(w\right)\log Q\left(w\right)}\nonumber \\
 & \qquad\times\e^{\left(1+s\right)n\sum_{w,x}T\left(w,x\right)\log Q\left(x|w\right)+\left(1+s\right)n\sum_{w,y}T\left(w,y\right)\log Q\left(y|w\right)-ns\sum_{x,y}T\left(x,y\right)\log\pi\left(x,y\right)}\label{eq:-8}\\
 & \leq\max_{T_{WXY}\in{\cal A}}\frac{1}{s}\left(H_{T}(XY|W)-D(T_{W}\|Q_{W})\right)+\frac{1+s}{s}\left(\sum_{w,x}T\left(w,x\right)\log Q\left(x|w\right)+\sum_{w,y}T\left(w,y\right)\log Q\left(y|w\right)\right)\nonumber \\
 & \qquad-\sum_{x,y}T\left(x,y\right)\log\pi\left(x,y\right)-\delta_{012,n}+\frac{1}{s}\delta_{n}\label{eq:-9-1}\\
 & \leq\sup_{P_{XY|W}\in\calA_{\epsilon}'}\left((1+\epsilon')\frac{1}{s}H(P_{XY|W}|Q_{W})-(1-\epsilon)\frac{1+s}{s}H_{Q}(XY|W)-\sum_{x,y}P\left(x,y\right)\log\pi\left(x,y\right)\right)\nonumber \\
 & \qquad-\delta_{012,n}+\frac{1}{s}\delta_{n},\label{eq:-38}
\end{align}

\hrulefill{}
\end{figure*}

Letting $n\rightarrow\infty$ in \eqref{eq:-38}, we have 
\begin{align}
 & \limsup_{n\to\infty}\frac{1}{n}D_{1+s}\left(P_{W^{n}X^{n}Y^{n}}\|P_{W^{n}}\pi_{X^{n}Y^{n}}\right)\nonumber \\
 & \leq\sup_{P_{XY|W}\in\calA_{\epsilon}'}\Bigl\{(1+\epsilon')\frac{1}{s}H(P_{XY|W}|Q_{W})\nonumber \\
 & \qquad-(1-\epsilon)\frac{1+s}{s}H_{Q}(XY|W)-\sum_{x,y}P\left(x,y\right)\log\pi\left(x,y\right)\Bigr\}.\label{eq:-15-1}
\end{align}
Since 1) $\epsilon>\epsilon'>0$ are arbitrary, 2) $H(P_{XY|W}|Q_{W})\leq\log\left\{ |\mathcal{X}||\mathcal{Y}|\right\} $,
and 3) $H_{Q}(X|W),H_{Q}(Y|W),\log\pi\left(x,y\right)$ are bounded
for $x,y\in\supp\left(\pi_{XY}\right)$, we have 
\begin{align}
 & \inf\left\{ R:D_{1+s}(P_{X^{n}Y^{n}|U_{n}}\|\pi_{X^{n}Y^{n}}|P_{U_{n}})\rightarrow0\right\} \nonumber \\
 & \leq\lim_{\epsilon\to0}\sup_{P_{XY|W}\in\calA_{\epsilon}'}\Bigl\{\frac{1}{s}H(P_{XY|W}|Q_{W})\nonumber \\
 & \qquad-(1-\epsilon)\frac{1+s}{s}H_{Q}(XY|W)-\sum_{x,y}P\left(x,y\right)\log\pi\left(x,y\right)\Bigr\}.\label{eq:-33}
\end{align}
Let $\left\{ \epsilon_{k}\right\} _{k=1}^{\infty}$ be a sequence
of decreasing positive numbers with $\lim_{k\to\infty}\epsilon_{k}=0$.
Assume $\left\{ P_{XY|W}^{\left(k\right)}\right\} _{k=1}^{\infty}$
is a sequence of optimal distributions $P_{XY|W}^{\left(k\right)}\in\calA_{\epsilon_{k}}'$
that attains the sup in \eqref{eq:-33} with $\epsilon$ there replaced
by $\epsilon_{k}$ (the sup is attained because we are optimizing
a continuous function over a compact set). Since $\mathcal{P}(\mathcal{X}\times\mathcal{Y}|\mathcal{W})$
is compact, there must exist some subsequence $P_{XY|W}^{\left(k_{i}\right)}$
that converges to some distribution $\widetilde{P}_{XY|W}$ as $i\to\infty$.
Since $\lim_{i\to\infty}\epsilon_{k_{i}}=0$, we must have 
\begin{align}
\widetilde{P}_{X|W} & =Q_{X|W}\textrm{ and }\widetilde{P}_{Y|W}=Q_{Y|W}.
\end{align}
Since $H(P_{XY|W}|Q_{W})$ and $\sum_{x,y}P\left(x,y\right)\log\pi\left(x,y\right)$
are continuous in $P_{XY|W}$, we have 
\begin{align}
 & \lim_{i\to\infty}\frac{1}{s}H(P_{XY|W}^{\left(k_{i}\right)}|Q_{W})-\sum_{x,y}P^{\left(k_{i}\right)}\left(x,y\right)\log\pi\left(x,y\right)\nonumber \\
 & =\frac{1}{s}H(\widetilde{P}_{XY|W}|Q_{W})-\sum_{x,y}\widetilde{P}\left(x,y\right)\log\pi\left(x,y\right).
\end{align}

Since the limit in \eqref{eq:-33} exists (by the monotonicity and
boundedness of the function in $\epsilon$), this limit must equal
the limit in \eqref{eq:-33} with $\epsilon$ replaced by the sequence
$\left\{ \epsilon_{k_{i}}\right\} _{i=1}^{\infty}$. Hence we obtain
that 
\begin{align}
 & \lim_{\epsilon\to0}\sup_{P_{XY|W}\in\calA_{\epsilon}'}\Bigl\{\frac{1}{s}H(P_{XY|W}|Q_{W})\nonumber \\
 & \quad-\frac{1+s}{s}H_{Q}(XY|W)-\sum_{x,y}P\left(x,y\right)\log\pi\left(x,y\right)\Bigr\}\nonumber \\
 & =\frac{1}{s}H(\widetilde{P}_{XY|W}|Q_{W})\nonumber \\
 & \qquad-\sum_{x,y}\widetilde{P}\left(x,y\right)\log\pi\left(x,y\right)-\frac{1+s}{s}H_{Q}(XY|W).\label{eq:-34}
\end{align}

Therefore, 
\begin{align}
 & \inf\left\{ R:D_{1+s}(P_{X^{n}Y^{n}|U_{n}}\|\pi_{X^{n}Y^{n}}|P_{U_{n}})\rightarrow0\right\} \nonumber \\
 & \leq\max_{\substack{P_{XY|W}:P_{X|W}=Q_{X|W},\\
P_{Y|W}=Q_{Y|W}
}
}\Bigl\{\frac{1}{s}H(P_{XY|W}|Q_{W})\nonumber \\
 & \qquad-\sum_{x,y}P\left(x,y\right)\log\pi\left(x,y\right)\Bigr\}-\frac{1+s}{s}H_{Q}(XY|W).\label{eq:-39}
\end{align}

Furthermore, since the distribution $Q_{WXY}$ is arbitrary, we can
minimize the bound above over all distributions satisfying $Q_{XY}=\pi_{XY}$
and $X-W-Y$. Hence 
\begin{align}
 & \inf\left\{ R:D_{1+s}(P_{X^{n}Y^{n}|U_{n}}\|\pi_{X^{n}Y^{n}}|P_{U_{n}})\rightarrow0\right\} \nonumber \\
 & \leq\Gamma_{1+s}^{\mathrm{UB}}(\pi_{XY}).
\end{align}
By the argument given at the end of the proof in \cite[Appendix A-A]{yu2018wyner},
the bound above is also an upper bound on the minimum rate for deterministic
codes.
\end{IEEEproof}

\section{\label{sec:Lower-Bound-for}Lower Bound for Case $s\in(0,\infty]$}

The proof in this section is similar to those of Theorems 1 and 2
in \cite{yu2018on}. By definition, we know that $\widetilde{T}_{1+s}(\pi_{XY})\leq T_{1+s}(\pi_{XY})$
and both of them are non-decreasing in $s$. On the other hand, $\widetilde{T}_{1}(\pi_{XY})=T_{1}(\pi_{XY})=C_{\mathsf{Wyner}}(X;Y)$.
Hence we have

\begin{equation}
C_{\mathsf{Wyner}}(X;Y)\leq\widetilde{T}_{1+s}(\pi_{XY})\leq T_{1+s}(\pi_{XY}),\;s\in(0,\infty].
\end{equation}

We next prove 
\begin{equation}
\widetilde{T}_{1+s}(\pi_{XY})\geq\Gamma_{1+s}^{\mathrm{LB}}(\pi_{XY}).
\end{equation}
The proof for this inequality is divided into three parts: Multi-letter
Expression for the Lower Bound, Single-letterization, Simplifying
Constraints.

\subsection{Multi-letter Expression for the Lower Bound}

To derive a multi-letter lower bound for $\widetilde{T}_{1+s}(\pi_{XY})$,
we need the following one-shot converse bound, which was proven in
\cite{yu2019renyi}.
\begin{lem}[One-Shot Bound for Converse Part]
\cite{yu2019renyi} \label{lem:oneshotcon} Assume $M\sim\mathrm{Unif}\{1,\ldots,\e^{R}\}$.
For any random mapping $P_{X|M}$, we define 
\begin{equation}
P_{MX}(m,x):=\e^{-R}P_{X|M}(x|m).\label{eq:-113-1-1}
\end{equation}
Then for $s\in[0,\infty]$ and any distribution $\pi_{X}$, we have
\begin{align}
 & D_{1+s}(P_{X}\|\pi_{X})\nonumber \\
 & \geq\max\left\{ D_{1+s}\left(P_{MX}\|P_{M}\pi_{X}\right)-R,D_{1+s}(P_{X}\|\pi_{X})\right\} .\label{eq:-146}
\end{align}
\end{lem}
By setting the tuple $\left(\pi_{X},P_{X|M},P_{M},R\right)$ to be
$\left(\pi_{XY}^{n},P_{X^{n}|M}P_{Y^{n}|M},P_{M},nR\right)$, Lemma
\ref{lem:oneshotcon} implies that 
\begin{align}
\widetilde{T}_{1+s}(\pi_{XY}) & \geq\inf_{\substack{\left\{ P_{M}P_{X^{n}|M}P_{Y^{n}|M}\right\} :\\
\frac{1}{n}D_{1+s}\left(P_{X^{n}Y^{n}}\|\pi_{XY}^{n}\right)\to0
}
}\nonumber \\
 & \qquad\limsup_{n\to\infty}\frac{1}{n}D_{1+s}\left(P_{MX^{n}Y^{n}}\|P_{M}\pi_{XY}^{n}\right).\label{eq:-146-1}
\end{align}
To lower bound the RHS of the inequality above, we need the following
lemma.
\begin{lem}
\label{lem:Assume-.-Then}Assume $P,Q\in\mathcal{P}(\mathcal{X})$.
Then for $s\in(0,\infty]$, we have 
\begin{align}
D_{1+s}(P\|Q) & =\sup_{R\in\mathcal{P}(\mathcal{X})}\frac{1}{s}\Bigl\{\sum_{x}R(x)\log P^{1+s}(x)Q^{-s}(x)\nonumber \\
 & \qquad-\sum_{x}R(x)\log R(x)\Bigr\}.\label{eq:-29}
\end{align}
\end{lem}
\begin{IEEEproof}
Observe that the objective function in the RHS of \eqref{eq:-29}
is concave in $R$. Define the Lagrangian function 
\begin{align}
\mathcal{L}(R,\lambda) & =\frac{1}{s}\Bigl\{\sum_{x}R(x)\log P^{1+s}(x)Q^{-s}(x)\nonumber \\
 & \qquad-\sum_{x}R(x)\log R(x)\Bigr\}+\lambda\left(\sum_{x}R(x)-1\right).
\end{align}
Hence letting the derivative of $\mathcal{L}(R,\lambda)$ respect
to $R(x)$ equal zero, we obtain that
\begin{align}
\frac{\partial\mathcal{L}(R,\lambda)}{\partial R(x)} & =\frac{1}{s}\left\{ \log P^{1+s}(x)Q^{-s}(x)-\left(1+\log R(x)\right)\right\} +\lambda\\
 & =0.
\end{align}
That is,
\begin{equation}
R(x)=\frac{P^{1+s}(x)Q^{-s}(x)}{\sum_{x}P^{1+s}(x)Q^{-s}(x)}.
\end{equation}
Hence 
\begin{align}
 & \textrm{RHS of }\eqref{eq:-29}\nonumber \\
 & =\frac{1}{s}\left\{ \sum_{x}R(x)\log P^{1+s}(x)Q^{-s}(x)-\sum_{x}R(x)\log R(x)\right\} \\
 & =D_{1+s}(P\|Q).
\end{align}
\end{IEEEproof}
By Lemma \ref{lem:Assume-.-Then} with $P\leftarrow P_{MX^{n}Y^{n}},Q\leftarrow P_{M}\pi_{XY}^{n},R\leftarrow P_{M}Q_{X^{n}Y^{n}|M}$
for $Q_{X^{n}Y^{n}|M}\in C(P_{X^{n}|M},P_{Y^{n}|M})$, we immediately
obtain \eqref{eq:-51}-\eqref{eq:-30} (given on the page \pageref{eq:-51}),
where in \eqref{eq:-30}, the $P(m)$'s in the logarithm have been
cancelled out. 
\begin{figure*}[t]
\begin{align}
 & D_{1+s}\left(P_{MX^{n}Y^{n}}\|P_{M}\pi_{XY}^{n}\right)\nonumber \\
 & \geq\max_{Q_{X^{n}Y^{n}|M}\in C(P_{X^{n}|M},P_{Y^{n}|M})}\frac{1}{s}\biggl\{\sum_{m,x^{n},y^{n}}P(m)Q(x^{n},y^{n}|m)\log\left[P^{1+s}(m,x^{n},y^{n})P^{-s}(m)\pi_{XY}^{n}(x^{n},y^{n})^{-s}\right]\nonumber \\
 & \qquad-\sum_{m,x^{n},y^{n}}P(m)Q(x^{n},y^{n}|m)\log\left[P(m)Q(x^{n},y^{n}|m)\right]\biggr\}\label{eq:-51}\\
 & =\max_{Q_{X^{n}Y^{n}|M}\in C(P_{X^{n}|M},P_{Y^{n}|M})}\left\{ \frac{1}{s}\sum_{m,x^{n},y^{n}}P(m)Q(x^{n},y^{n}|m)\log\frac{1}{\pi_{XY}^{n}(x^{n},y^{n})^{s}Q(x^{n},y^{n}|m)}\right\} -\frac{1+s}{s}H(X^{n}Y^{n}|M).\label{eq:-30}
\end{align}

\hrulefill{}
\end{figure*}

Therefore, we obtain the multi-letter lower bound given in \eqref{eq:-18}
(given on the page \pageref{eq:-18}).

\begin{figure*}[t]
\begin{align}
\widetilde{T}_{1+s}(\pi_{XY}) & \geq\frac{1}{n}\inf_{\substack{\left\{ P_{M}P_{X^{n}|M}P_{Y^{n}|M}\right\} :\\
\frac{1}{n}D_{1+s}\left(P_{X^{n}Y^{n}}\|\pi_{XY}^{n}\right)\to0
}
}\limsup_{n\to\infty}-\frac{1+s}{s}H(X^{n}Y^{n}|M)\nonumber \\
 & \qquad+\frac{1}{s}\max_{Q_{X^{n}Y^{n}|M}\in C(P_{X^{n}|M},P_{Y^{n}|M})}\left\{ \sum_{m,x^{n},y^{n}}P(m)Q(x^{n},y^{n}|m)\log\frac{1}{\pi_{XY}^{n}(x^{n},y^{n})^{s}Q(x^{n},y^{n}|m)}\right\} .\label{eq:-18}
\end{align}

\hrulefill{}
\end{figure*}

\subsection{Single-letterization}

Next we derive a single-letter lower bound for the RHS of \eqref{eq:-18}.
 Denote $J\sim P_{J}:=\mathrm{Unif}[1:n]$ as a time index independent
of $(M,X^{n},Y^{n})$. Then 
\begin{align}
 & -\frac{1}{n}\sum_{i=1}^{n}H(X_{i}|X^{i-1}M)-\frac{1}{n}\sum_{i=1}^{n}H(Y_{i}|Y^{i-1}M)\nonumber \\
 & =-H(X_{J}|X^{J-1}MJ)-H(Y_{J}|Y^{J-1}MJ).\label{eq:-21}
\end{align}

Next we single-letterize the last term in \eqref{eq:-18}. Observe
that 
\begin{align}
 & \sum_{x^{n},y^{n}}Q(x^{n},y^{n}|m)\log\frac{1}{\pi_{XY}^{n}(x^{n},y^{n})^{s}Q(x^{n},y^{n}|m)}\nonumber \\
 & =\sum_{i=1}^{n}\sum_{x_{i},y_{i}}\sum_{x^{i-1},y^{i-1}}Q(x^{i-1},y^{i-1}|m)Q(x_{i},y_{i}|x^{i-1},y^{i-1},m)\nonumber \\
 & \qquad\times\log\frac{1}{\pi\left(x_{i},y_{i}\right)^{s}Q(x_{i},y_{i}|x^{i-1},y^{i-1},m)}\\
 & \geq\sum_{i=1}^{n}\min_{\substack{\widetilde{Q}_{X^{i-1}Y^{i-1}|M}\in\\
C(P_{X^{i-1}|M},P_{Y^{i-1}|M})
}
}\sum_{x_{i},y_{i}}\sum_{x^{i-1},y^{i-1}}\widetilde{Q}(x^{i-1},y^{i-1}|m)\nonumber \\
 & \qquad\times Q(x_{i},y_{i}|x^{i-1},y^{i-1},m)\nonumber \\
 & \qquad\times\log\frac{1}{\pi\left(x_{i},y_{i}\right)^{s}Q(x_{i},y_{i}|x^{i-1},y^{i-1},m)}.\label{eq:-19}
\end{align}
Now we employ the following ``chain rule'' on coupling sets. Since
the following lemma is easy to verify,  we omit its proof.
\begin{lem}[Chain Rule on Coupling Sets]
\label{lem:coupling} For a pair of conditional distributions $(P_{X^{n}|W},P_{Y^{n}|W})$,
we have
\begin{equation}
\prod_{i=1}^{n}C(P_{X_{i}|X^{i-1}W},P_{Y_{i}|Y^{i-1}W})\subseteq C(P_{X^{n}|W},P_{Y^{n}|W}),
\end{equation}
where for $i\in[1:n]$,
\begin{align}
 & C(P_{X_{i}|X^{i-1}W},P_{Y_{i}|Y^{i-1}W})\nonumber \\
 & :=\Bigl\{ Q_{X_{i}Y_{i}|X^{i-1}Y^{i-1}W}:\,Q_{X_{i}|X^{i-1}Y^{i-1}W}=P_{X_{i}|X^{i-1}W},\nonumber \\
 & \qquad\:Q_{Y_{i}|X^{i-1}Y^{i-1}W}=P_{Y_{i}|Y^{i-1}W}\Bigr\}
\end{align}
and
\begin{align}
 & \prod_{i=1}^{n}C(P_{X_{i}|X^{i-1}W},P_{Y_{i}|Y^{i-1}W})\nonumber \\
 & :=\Bigl\{\prod_{i=1}^{n}Q_{X_{i}Y_{i}|X^{i-1}Y^{i-1}W}:\,Q_{X_{i}Y_{i}|X^{i-1}Y^{i-1}W}\nonumber \\
 & \qquad\in C(P_{X_{i}|X^{i-1}W},P_{Y_{i}|Y^{i-1}W}),\forall i\in[1:n]\Bigr\}.
\end{align}
\end{lem}
By Lemma \ref{lem:coupling}, we have that for any function $f:\mathcal{P}\left(\mathcal{X}^{n}\times\mathcal{Y}^{n}\right)\to\mathbb{R}$,
\begin{align}
 & \max_{Q_{X^{n}Y^{n}|W}\in C(P_{X^{n}|W},P_{Y^{n}|W})}f\left(Q_{X^{n}Y^{n}|W}\right)\nonumber \\
 & \geq\max_{\substack{Q_{X^{n}Y^{n}|W}\in\\
\prod_{i=1}^{n}C(P_{X_{i}|X^{i-1}W},P_{Y_{i}|Y^{i-1}W})
}
}f\left(\prod_{i=1}^{n}Q_{X_{i}Y_{i}|X^{i-1}Y^{i-1}W}\right).\label{eq:-57}
\end{align}
Therefore, substituting \eqref{eq:-19} into the last term in \eqref{eq:-18}
and utilizing \eqref{eq:-57}, we obtain \eqref{eq:-52}-\eqref{eq:-36}
(given on the page \pageref{eq:-52}). Here \eqref{eq:-22} follows
from \eqref{eq:-57}. The swapping of min and max in \eqref{eq:-14}
follows since on one hand, minimax is no smaller than maximin, and
on the other hand, 
\begin{align}
\eqref{eq:-22} & \geq\sum_{m}P(m)\sum_{i=1}^{n}\min_{\substack{\widetilde{Q}_{X^{i-1}Y^{i-1}|M}\in\\
C(P_{X^{i-1}|M},P_{Y^{i-1}|M})
}
}\nonumber \\
 & \qquad\sum_{x^{i-1},y^{i-1}}\widetilde{Q}(x^{i-1},y^{i-1}|m)\nonumber \\
 & \qquad\times\sum_{x_{i},y_{i}}Q^{*}(x_{i},y_{i}|x^{i-1},y^{i-1},m)\nonumber \\
 & \qquad\times\log\frac{1}{\pi\left(x_{i},y_{i}\right)^{s}Q^{*}(x_{i},y_{i}|x^{i-1},y^{i-1},m)}\label{eq:-15-2}\\
 & =\eqref{eq:-14}
\end{align}
with 
\begin{align}
 & Q_{X_{i}Y_{i}|X^{i-1}Y^{i-1}M}^{*}\nonumber \\
 & :=\arg\max_{\substack{Q_{X_{i}Y_{i}|X^{i-1}Y^{i-1}M}\in\\
C(P_{X_{i}|X^{i-1}M},P_{Y_{i}|Y^{i-1}M})
}
}\sum_{x_{i},y_{i}}Q(x_{i},y_{i}|x^{i-1},y^{i-1},m)\nonumber \\
 & \qquad\times\log\frac{1}{\pi\left(x_{i},y_{i}\right)^{s}Q(x_{i},y_{i}|x^{i-1},y^{i-1},m)}.
\end{align}
\begin{figure*}[t]
\begin{align}
 & \max_{\substack{Q_{X^{n}Y^{n}|M}\in\\
C(P_{X^{n}|M},P_{Y^{n}|M})
}
}\left\{ \sum_{m,x^{n},y^{n}}P(m)Q(x^{n},y^{n}|m)\log\frac{1}{\pi_{XY}^{n}(x^{n},y^{n})^{s}Q(x^{n},y^{n}|m)}\right\} \nonumber \\
 & \geq\sum_{m}P(m)\max_{\substack{Q_{X^{n}Y^{n}|M}\in\\
C(P_{X^{n}|M},P_{Y^{n}|M})
}
}\sum_{i=1}^{n}\min_{\substack{\widetilde{Q}_{X^{i-1}Y^{i-1}|M}\in\\
C(P_{X^{i-1}|M},P_{Y^{i-1}|M})
}
}\sum_{x_{i},y_{i}}\sum_{x^{i-1},y^{i-1}}\widetilde{Q}(x^{i-1},y^{i-1}|m)Q(x_{i},y_{i}|x^{i-1},y^{i-1},m)\nonumber \\
 & \qquad\times\log\frac{1}{\pi\left(x_{i},y_{i}\right)^{s}Q(x_{i},y_{i}|x^{i-1},y^{i-1},m)}\label{eq:-52}\\
 & \geq\sum_{m}P(m)\sum_{i=1}^{n}\max_{\substack{Q_{X_{i}Y_{i}|X^{i-1}Y^{i-1}M}\in\\
C(P_{X_{i}|X^{i-1}M},P_{Y_{i}|Y^{i-1}M})
}
}\min_{\substack{\widetilde{Q}_{X^{i-1}Y^{i-1}|M}\in\\
C(P_{X^{i-1}|M},P_{Y^{i-1}|M})
}
}\sum_{x_{i},y_{i}}\sum_{x^{i-1},y^{i-1}}\widetilde{Q}(x^{i-1},y^{i-1}|m)Q(x_{i},y_{i}|x^{i-1},y^{i-1},m)\nonumber \\
 & \qquad\times\log\frac{1}{\pi\left(x_{i},y_{i}\right)^{s}Q(x_{i},y_{i}|x^{i-1},y^{i-1},m)}\label{eq:-22}\\
 & =\sum_{m}P(m)\sum_{i=1}^{n}\min_{\substack{\widetilde{Q}_{X^{i-1}Y^{i-1}|M}\in\\
C(P_{X^{i-1}|M},P_{Y^{i-1}|M})
}
}\sum_{x^{i-1},y^{i-1}}\widetilde{Q}(x^{i-1},y^{i-1}|m)\max_{\substack{Q_{X_{i}Y_{i}|X^{i-1}Y^{i-1}M}\in\\
C(P_{X_{i}|X^{i-1}M},P_{Y_{i}|Y^{i-1}M})
}
}\sum_{x_{i},y_{i}}Q(x_{i},y_{i}|x^{i-1},y^{i-1},m)\nonumber \\
 & \qquad\times\log\frac{1}{\pi\left(x_{i},y_{i}\right)^{s}Q(x_{i},y_{i}|x^{i-1},y^{i-1},m)}\label{eq:-14}\\
 & =n\sum_{m}P(m)\sum_{i=1}^{n}P_{J}(i)\min_{\substack{\widetilde{Q}_{X^{J-1}Y^{J-1}|MJ}\in\\
C(P_{X^{J-1}|MJ},P_{Y^{J-1}|MJ})
}
}\sum_{x^{i-1},y^{i-1}}\widetilde{Q}_{X^{J-1}Y^{J-1}|MJ}(x^{i-1},y^{i-1}|m,i)\max_{\substack{Q_{X_{J}Y_{J}|X^{J-1}Y^{J-1}MJ}\in\\
C(P_{X_{J}|X^{J-1}MJ},P_{Y_{J}|Y^{J-1}MJ})
}
}\nonumber \\
 & \qquad\sum_{x,y}Q_{X_{J}Y_{J}|X^{J-1}Y^{J-1}MJ}(x,y|x^{i-1},y^{i-1},m,i)\log\frac{1}{\pi\left(x,y\right)^{s}Q_{X_{J}Y_{J}|X^{J-1}Y^{J-1}MJ}(x,y|x^{i-1},y^{i-1},m,i)}.\label{eq:-36}
\end{align}

\hrulefill{}
\end{figure*}

Denote 
\begin{equation}
W:=MJ,U:=X^{J-1},V:=Y^{J-1},X:=X_{J},Y:=Y_{J}.\label{eq:-37}
\end{equation}
  It is easy to verify that for $s>0$, $\frac{1}{n}D_{1+s}\left(P_{X^{n}Y^{n}}\|\pi_{XY}^{n}\right)\le\epsilon$
implies $D\left(P_{XY}\|\pi_{XY}\right)\le\epsilon$. Since $\pi_{XY}$
has a finite support, $D\left(P_{XY}\|\pi_{XY}\right)\to0$ if and
only if $D_{\infty}\left(P_{XY}\|\pi_{XY}\right)\to0$. Therefore,
substituting \eqref{eq:-21} and \eqref{eq:-36} into \eqref{eq:-18}
and utilizing the identification of the random variables in \eqref{eq:-37},
we obtain \eqref{eq:-40} (given on the page \pageref{eq:-40}). 
\begin{figure*}[t]
\begin{align}
\widetilde{T}_{1+s}(\pi_{XY}) & \geq\lim_{\epsilon\downarrow0}\inf_{\substack{P_{W}P_{U|W}P_{V|W}P_{X|UW}P_{Y|VW}:\\
D_{\infty}\left(P_{XY}\|\pi_{XY}\right)\le\epsilon
}
}-\frac{1+s}{s}\left(H(X|UW)+H(Y|VW)\right)\nonumber \\
 & \quad+\frac{1}{s}\sum_{w}P(w)\inf_{\substack{\widetilde{Q}_{UV|W}\in\\
C(P_{U|W},P_{V|W})
}
}\sum_{u,v}\widetilde{Q}(u,v|w)\max_{\substack{Q_{XY|UVW}\in\\
C(P_{X|UW},P_{Y|VW})
}
}\sum_{x,y}Q(x,y|u,v,w)\log\frac{1}{\pi\left(x,y\right)^{s}Q(x,y|u,v,w)}.\label{eq:-40}
\end{align}

\hrulefill{}
\end{figure*}
 For $\widetilde{Q}_{UV|W}\in C(P_{U|W},P_{V|W})$, define the following
induced joint distribution as 
\begin{align}
 & \widehat{Q}_{\left(U,V',W\right),\left(U',V,W'\right)}(u,v',w,u',v,w')\nonumber \\
 & :=P_{W}(w)\widetilde{Q}_{UV|W}(u,v|w)1\left\{ w'=w\right\} \nonumber \\
 & \qquad\times P_{V|W}(v'|w)P_{U|W}(u'|w').
\end{align}
Then this joint distribution satisfies the following marginal constraints:
\begin{align}
\widehat{Q}_{UVW}(u,v,w) & =P_{W}(w)\widetilde{Q}_{UV|W}(u,v|w)\label{eq:-47}\\
\widehat{Q}_{UV'W}(u,v',w) & =P_{UVW}(u,v',w)\\
\widehat{Q}_{U'VW'}(u',v,w') & =P_{UVW}(u',v,w').\label{eq:-48}
\end{align}
Utilizing this induced distribution, its properties in \eqref{eq:-47}-\eqref{eq:-48},
and the lower bound in \eqref{eq:-40}, we obtain \eqref{eq:-53}-\eqref{eq:-49}
(given on the page \pageref{eq:-53}). 
\begin{figure*}[t]
\begin{align}
\widetilde{T}_{1+s}(\pi_{XY}) & \geq\lim_{\epsilon\downarrow0}\inf_{\substack{P_{W}P_{U|W}P_{V|W}P_{X|UW}P_{Y|VW}:\\
D_{\infty}\left(P_{XY}\|\pi_{XY}\right)\le\epsilon
}
}-\frac{1+s}{s}\left(H(X|UW)+H(Y|VW)\right)\nonumber \\
 & \quad+\frac{1}{s}\inf_{\substack{\widetilde{Q}_{UV|W}\in\\
C(P_{U|W},P_{V|W})
}
}\sum_{u,u',v,v',w,w'}\widehat{Q}(u,v',w,u',v,w')\max_{\substack{Q_{XY}\in\\
C(P_{X|UW=u,w},P_{Y|VW=v,w'})
}
}\sum_{x,y}Q(x,y)\log\frac{1}{\pi\left(x,y\right)^{s}Q(x,y)}\label{eq:-53}\\
 & \geq\lim_{\epsilon\downarrow0}\inf_{\substack{P_{W}P_{U|W}P_{V|W}P_{X|UW}P_{Y|VW}:\\
D_{\infty}\left(P_{XY}\|\pi_{XY}\right)\le\epsilon
}
}-\frac{1+s}{s}\left(H(X|UW)+H(Y|VW)\right)\nonumber \\
 & \quad+\frac{1}{s}\inf_{\substack{\widehat{Q}_{\left(U,V',W\right),\left(U',V,W'\right)}\in\\
C(P_{UVW},P_{UVW})
}
}\sum_{u,u',v,v',w,w'}\widehat{Q}(u,v',w,u',v,w')\nonumber \\
 & \quad\times\max_{\substack{Q_{XY}\in\\
C(P_{X|\left(U,V,W\right)=\left(u,v',w\right)},P_{Y|\left(U,V,W\right)=\left(u',v,w'\right)})
}
}\sum_{x,y}Q(x,y)\log\frac{1}{\pi\left(x,y\right)^{s}Q(x,y)}.\label{eq:-49}
\end{align}

\hrulefill{}
\end{figure*}
Substituting $W\leftarrow\left(U,V,W\right)$, we can simplify \eqref{eq:-49}
as \eqref{eq:-23} (given on the page \pageref{eq:-23}). 
\begin{figure*}[t]
\begin{align}
\widetilde{T}_{1+s}(\pi_{XY}) & \geq\lim_{\epsilon\downarrow0}\inf_{P_{W}P_{X|W}P_{Y|W}:D_{\infty}\left(P_{XY}\|\pi_{XY}\right)\le\epsilon}-\frac{1+s}{s}\left(H(X|W)+H(Y|W)\right)\nonumber \\
 & \qquad+\frac{1}{s}\inf_{Q_{WW'}\in C(P_{W},P_{W})}\sum_{w,w'}Q_{WW'}(w,w')\max_{Q_{XY}\in C(P_{X|W=w},P_{Y|W=w'})}\sum_{x,y}Q(x,y)\log\frac{1}{\pi\left(x,y\right)^{s}Q(x,y)}.\label{eq:-23}
\end{align}

\hrulefill{}
\end{figure*}

\subsection{Simplifying Constraints}

Next we prove that the constraint $D_{\infty}\left(P_{XY}\|\pi_{XY}\right)\le\epsilon$
in \eqref{eq:-23} can be replaced by $P_{XY}=\pi_{XY}$. For two
distributions $\left(P_{XY},\pi_{XY}\right)$ such that $D_{\infty}\left(P_{XY}\|\pi_{XY}\right)\le\epsilon$,
we can write
\begin{equation}
\pi_{XY}\left(x,y\right)=e^{-\epsilon}P_{XY}\left(x,y\right)+\left(1-e^{-\epsilon}\right)\widehat{P}_{XY}\left(x,y\right),\label{eq:-100-2}
\end{equation}
where 
\begin{equation}
\widehat{P}_{XY}\left(x,y\right):=\frac{e^{\epsilon}\pi_{XY}\left(x,y\right)-P_{XY}\left(x,y\right)}{e^{\epsilon}-1}.
\end{equation}
Note that $\supp\left(\widehat{P}_{XY}\right)\subseteq\supp\left(\pi_{XY}\right).$
Define 
\begin{align}
 & \widetilde{P}_{XYWU}(x,y,w,u)\nonumber \\
 & =\begin{cases}
e^{-\epsilon}P_{W}(w)P_{X|W}(x|w)P_{Y|W}(y|w) & \textrm{if }u=1\\
\left(1-e^{-\epsilon}\right)\widehat{P}_{WXY}\left(w,x,y\right) & \textrm{if }u=0
\end{cases},\label{eq:-43}
\end{align}
where $\widehat{P}_{WXY}\left(w,x,y\right):=\widehat{P}_{XY}\left(x,y\right)1\left\{ w=(x,y)\right\} $.
Obviously, 
\begin{equation}
X\to(U,W)\to Y\label{eq:-45}
\end{equation}
forms a Markov chain under the distribution $\widetilde{P}$, and
moreover, 
\begin{align}
\widetilde{P}_{XY} & =\pi_{XY}.\label{eq:-17}
\end{align}
Now consider the expression in \eqref{eq:-50} (on the page \pageref{eq:-50})
induced by $\left(\widetilde{P}_{XYWU},\pi_{XY}\right)$. Then we
have the following upper bound on $\varphi_{s}\left(\widetilde{P}_{XYWU},\pi_{XY}\right)$.
\begin{figure*}[t]
\begin{align}
\varphi_{s}\left(\widetilde{P}_{XYWU},\pi_{XY}\right) & :=-\frac{1+s}{s}H_{\widetilde{P}}(XY|WU)+\frac{1}{s}\inf_{Q_{WUW'U'}\in C(\widetilde{P}_{WU},\widetilde{P}_{WU})}\sum_{w,u,w',u'}Q(w,u,w',u')\nonumber \\
 & \qquad\times\max_{\substack{Q_{XY}\in\\
C(\widetilde{P}_{X|(W,U)=(w,u)},\widetilde{P}_{Y|(W,U)=(w',u')})
}
}\sum_{x,y}Q(x,y)\log\frac{1}{\pi\left(x,y\right)^{s}Q(x,y)}.\label{eq:-50}
\end{align}

\hrulefill{}
\end{figure*}

\begin{lem}
Given the definition of $\widetilde{P}_{XYWU}$ in \eqref{eq:-43},
we have 
\begin{align}
 & \varphi_{s}\left(\widetilde{P}_{XYWU},\pi_{XY}\right)\nonumber \\
 & \leq e^{-\epsilon}\biggl(-\frac{1+s}{s}H(XY|W)+\frac{1}{s}\inf_{\substack{Q_{WW'}\in\\
C(P_{W},P_{W})
}
}\sum_{w,w'}Q(w,w')\nonumber \\
 & \qquad\times\max_{\substack{Q_{XY}\in\\
C(P_{X|W=w},P_{Y|W=w'})
}
}\sum_{x,y}Q(x,y)\log\frac{1}{\pi\left(x,y\right)^{s}Q(x,y)}\biggr)\nonumber \\
 & \qquad+o(\epsilon),\label{eq:-44}
\end{align}
where 
\begin{equation}
o(\epsilon)=\left(1-e^{-\epsilon}\right)\max_{\left(x,y\right)\in\supp(\pi_{XY})}\log\frac{1}{\pi\left(x,y\right)}
\end{equation}
 vanishes as $\epsilon\downarrow0$.
\end{lem}
\begin{IEEEproof}
Starting from the definition of $\varphi_{s}\left(\widetilde{P}_{XYWU},\pi_{XY}\right)$
in \eqref{eq:-50}, we have \eqref{eq:-13}-\eqref{eq:-15} (given
on the page \pageref{eq:-13}), where \eqref{eq:-13} follows since
$H_{\widetilde{P}}(XY|WU)\geq e^{-\epsilon}H(XY|W)$ (from the definition
of $\widetilde{P}_{XYWU}$ in \eqref{eq:-43}) and 
\begin{align}
 & \widetilde{P}_{U}(u)1\left\{ u'=u\right\} Q_{WW'|\left(U,U'\right)=\left(u,u'\right)}(w,w')\nonumber \\
 & \in C(\widetilde{P}_{WU},\widetilde{P}_{WU})
\end{align}
for any $Q_{WW'|\left(U,U'\right)=\left(u,u'\right)}\in C(\widetilde{P}_{W|U=u},\widetilde{P}_{W|U=u'})$;
and \eqref{eq:-15} follows from the definition of $\widetilde{P}_{XYWU}$
in \eqref{eq:-43}. Observe that if we set $Q_{WW'}(w,w')=\widehat{P}_{W}(w)1\{w'=w\}$
in the last term in \eqref{eq:-15}, then this term would be equal
to $\left(1-e^{-\epsilon}\right)\sum_{x,y}\widehat{P}_{XY}(x,y)\log\frac{1}{\pi\left(x,y\right)}$
which is no larger than $\left(1-e^{-\epsilon}\right)\max_{\left(x,y\right)\in\supp(\pi_{XY})}\log\frac{1}{\pi\left(x,y\right)}=o(\epsilon)$
(since $\supp\left(\widehat{P}_{XY}\right)\subseteq\supp\left(\pi_{XY}\right)$).
Hence we have inequality \eqref{eq:-44}.
\begin{figure*}[t]
\begin{align}
 & \varphi_{s}\left(\widetilde{P}_{XYWU},\pi_{XY}\right)\nonumber \\
 & \leq-e^{-\epsilon}\frac{1+s}{s}H(XY|W)+\frac{1}{s}\sum_{u,u'}\widetilde{P}_{U}(u)1\left\{ u'=u\right\} \inf_{Q_{WW'}\in C(\widetilde{P}_{W|U=u},\widetilde{P}_{W|U=u'})}\sum_{w,w'}Q(w,w')\nonumber \\
 & \qquad\times\max_{Q_{XY}\in C(\widetilde{P}_{X|(W,U)=(w,u)},\widetilde{P}_{Y|(W,U)=(w',u')})}\sum_{x,y}Q(x,y)\log\frac{1}{\pi\left(x,y\right)^{s}Q(x,y)}\label{eq:-13}\\
 & =-e^{-\epsilon}\frac{1+s}{s}H(XY|W)+e^{-\epsilon}\frac{1}{s}\inf_{Q_{WW'}\in C(P_{W},P_{W})}\sum_{w,w'}Q(w,w')\max_{Q_{XY}\in C(P_{X|W=w},P_{Y|W=w'})}\sum_{x,y}Q(x,y)\log\frac{1}{\pi\left(x,y\right)^{s}Q(x,y)}\nonumber \\
 & \qquad+\left(1-e^{-\epsilon}\right)\frac{1}{s}\inf_{Q_{WW'}\in C(\widehat{P}_{W},\widehat{P}_{W})}\sum_{w,w'}Q(w,w')\max_{Q_{XY}\in C(\widehat{P}_{X|W=w},\widehat{P}_{Y|W=w'})}\sum_{x,y}Q(x,y)\log\frac{1}{\pi\left(x,y\right)^{s}Q(x,y)}.\label{eq:-15}
\end{align}

\hrulefill{}
\end{figure*}
\end{IEEEproof}
Using \eqref{eq:-44} and the lower bound in \eqref{eq:-23}, we obtain
that  
\begin{align}
 & \widetilde{T}_{1+s}(\pi_{XY})\nonumber \\
 & \geq\lim_{\epsilon\downarrow0}\inf_{\substack{P_{W}P_{X|W}P_{Y|W}:\\
D_{\infty}\left(P_{XY}\|\pi_{XY}\right)\le\epsilon
}
}e^{\epsilon}\left(\varphi_{s}\left(\widetilde{P}_{XYWU},\pi_{XY}\right)+o(\epsilon)\right)\\
 & =\lim_{\epsilon\downarrow0}\inf_{\substack{P_{W}P_{X|W}P_{Y|W}:\\
D_{\infty}\left(P_{XY}\|\pi_{XY}\right)\le\epsilon
}
}\varphi_{s}\left(\widetilde{P}_{XYWU},\pi_{XY}\right)\label{eq:-16}\\
 & \geq\inf_{\substack{\widetilde{P}_{WU}\widetilde{P}_{X|WU}\widetilde{P}_{Y|WU}:\\
\widetilde{P}_{XY}=\pi_{XY}
}
}\varphi_{s}\left(\widetilde{P}_{XYWU},\pi_{XY}\right)\label{eq:-12}\\
 & =\Gamma_{1+s}^{\mathrm{LB}}(\pi_{XY}).
\end{align}
where \eqref{eq:-16} follows since $e^{\epsilon}$ and any implied
constants in the notation $o(\epsilon)$ do not depend on $P_{W}P_{X|W}P_{Y|W}$,
and \eqref{eq:-12} follows since the distribution $\widetilde{P}_{XYWU}$
in \eqref{eq:-16} satisfies $X\to(U,W)\to Y$ and $\widetilde{P}_{XY}=\pi_{XY}$
(see \eqref{eq:-45} and \eqref{eq:-17}).

\appendices{}

\section{\label{sec:Proof-of-Lemma-property}Proof of Lemma \ref{lem:property}}

1) Observe that 
\begin{align*}
 & -\frac{1+s}{s}H(XY|W)\\
 & \qquad+\sum_{w}P(w)\mathcal{H}_{s}(P_{X|W=w},P_{Y|W=w}\|\pi_{XY})
\end{align*}
is a linear function of $P_{W}$. Hence Statement 1) can be proven
by standard cardinality bounding techniques (e.g., the support lemma
in \cite[Appendix C]{Gamal}).

2) Observe that $\Gamma_{1+s}^{\mathrm{UB}}(\pi_{XY})$ can be rewritten
as
\begin{align}
 & \Gamma_{1+s}^{\mathrm{UB}}(\pi_{XY})\nonumber \\
 & =\min_{\substack{P_{W}P_{X|W}P_{Y|W}:\\
P_{XY}=\pi_{XY}
}
}\max_{\substack{Q_{XY|W}\in\\
C(P_{X|W},P_{Y|W})
}
}\psi_{s}\left(P_{W}P_{X|W}P_{Y|W},Q_{XY|W}\right),\label{eq:}
\end{align}
where 
\begin{align}
 & \psi_{s}\left(P_{W}P_{X|W}P_{Y|W},Q_{XY|W}\right)\nonumber \\
 & :=\sum_{w,x,y}P(w)Q(x,y|w)\log\frac{P(x|w)P(y|w)}{P\left(x,y\right)}\nonumber \\
 & \qquad+\frac{1}{s}\left(H(Q_{XY|W}|P_{W})-H(XY|W)\right).\label{eq:-35}
\end{align}
On the other hand, under the constraint $Q_{XY|W}\in C(P_{X|W},P_{Y|W})$,
\begin{align}
H(Q_{XY|W}|P_{W}) & \le H(Q_{X|W}|P_{W})+H(Q_{Y|W}|P_{W})\\
 & =H(XY|W).\label{eq:-4}
\end{align}
Hence for any $0<s<s'$, 
\begin{align}
 & \frac{1}{s}\left(H(Q_{XY|W}|P_{W})-H(XY|W)\right)\nonumber \\
 & \leq\frac{1}{s'}\left(H(Q_{XY|W}|P_{W})-H(XY|W)\right).\label{eq:-1}
\end{align}
Combining \eqref{eq:} and \eqref{eq:-1}, we obtain that $\Gamma_{1+s}^{\mathrm{UB}}(\pi_{XY})\le\Gamma_{1+s'}^{\mathrm{UB}}(\pi_{XY})$,
i.e., $\Gamma_{1+s}^{\mathrm{UB}}(\pi_{XY})$ is non-decreasing in
$s\in(0,\infty)$.

Now we consider $\Gamma_{1+s}^{\mathrm{LB}}(\pi_{XY})$. Following
similar steps above, one can obtain that $\Gamma_{1+s}^{\mathrm{LB}}(\pi_{XY})$
is non-decreasing in $s\in(0,\infty)$.

3) For distribution $P_{W}P_{X|W}P_{Y|W}$ such that $P_{XY}=\pi_{XY}$
and distribution $Q_{XY|W}\in C(P_{X|W},P_{Y|W})$, we have that 
\begin{equation}
-H(\pi_{XY})\leq H(Q_{XY|W}|P_{W})-H(XY|W)\leq0,\label{eq:-31}
\end{equation}
where the first inequality above follows since $H(Q_{XY|W}|P_{W})\ge0$
and $H(XY|W)\le H(\pi_{XY})$; and the second inequality follows by
\eqref{eq:-4}.

Define 
\begin{align}
\Gamma^{\mathrm{UB}}(\pi_{XY}) & :=\min_{\substack{P_{W}P_{X|W}P_{Y|W}:\\
P_{XY}=\pi_{XY}
}
}\max_{\substack{Q_{XY|W}\in\\
C(P_{X|W},P_{Y|W})
}
}\sum_{w,x,y}P(w)\nonumber \\
 & \qquad\times Q(x,y|w)\log\frac{P(x|w)P(y|w)}{\pi\left(x,y\right)}.\label{eq:-2}
\end{align}
Hence by combining \eqref{eq:} and \eqref{eq:-31}, for $s\in(0,\infty)$,
\begin{equation}
\Gamma^{\mathrm{UB}}(\pi_{XY})-\frac{1}{s}H(\pi_{XY})\leq\Gamma_{1+s}^{\mathrm{UB}}(\pi_{XY})\leq\Gamma^{\mathrm{UB}}(\pi_{XY}).
\end{equation}
Letting $s\to\infty$, we obtain $\Gamma_{\infty}^{\mathrm{UB}}(\pi_{XY})=\Gamma^{\mathrm{UB}}(\pi_{XY})$,
i.e., equality \eqref{eq:-36-5}. Equality \eqref{eq:-41} can be
proven similarly.

By choosing $Q_{XY|W}=P_{X|W}P_{Y|W}$ in \eqref{eq:}, we obtain
for $s\in(0,\infty)$, 
\begin{equation}
\Gamma_{1+s}^{\mathrm{UB}}(\pi_{XY})\geq C_{\mathsf{Wyner}}(X;Y).\label{eq:-6}
\end{equation}
Let $\left\{ s_{k}\right\} _{k=1}^{\infty}$ be a sequence of decreasing
positive numbers with $\lim_{k\to\infty}s_{k}=0$. Assume that $(P_{W},P_{X|W},P_{Y|W})$
attains $C_{\mathsf{Wyner}}(X;Y)$. For this optimal $(P_{W},P_{X|W},P_{Y|W})$,
assume that $\left\{ Q_{XY|W}^{(k)}\right\} _{k=1}^{\infty}$ is a
sequence of optimal distributions in which $Q_{XY|W}^{(k)}$ attains
the maximum in the following optimization:
\begin{equation}
\gamma_{k}:=\max_{Q_{XY|W}\in C(P_{X|W},P_{Y|W})}\psi_{s_{k}}\left(P_{W}P_{X|W}P_{Y|W},Q_{XY|W}\right)\label{eq:-11}
\end{equation}
where $\psi_{s}\left(\cdot\right)$ is defined in \eqref{eq:-35}.
Since the space $\mathcal{P}(\mathcal{X}\times\mathcal{Y}|\mathcal{W})$
of finitely-supported conditional distributions $Q_{XY|W}$ is compact,
there exists some subsequence $Q_{XY|W}^{\left(k_{i}\right)}$ that
converges to some distribution $\widetilde{Q}_{XY|W}$ as $i\to\infty$.
Since $Q_{XY|W}^{\left(k_{i}\right)}\in C(P_{X|W},P_{Y|W})$ and $H(Q_{XY|W}^{\left(k_{i}\right)}|P_{W})\leq H(P_{X|W}P_{Y|W}|P_{W})$
(see \eqref{eq:-4}), we must have 
\begin{align}
\widetilde{Q}_{X|W} & =P_{X|W}\\
\widetilde{Q}_{Y|W} & =P_{Y|W}\\
H(\widetilde{Q}_{XY|W}|P_{W}) & \leq H(P_{X|W}P_{Y|W}|P_{W}).\label{eq:-10}
\end{align}
Now we claim that equality holds in \eqref{eq:-10}. Suppose, to the
contrary, that the inequality in \eqref{eq:-10} is strict. Then observe
that 
\begin{align}
\gamma_{k_{i}} & =\max_{Q_{XY|W}\in C(P_{X|W},P_{Y|W})}\psi_{s_{k_{i}}}\left(P_{W}P_{X|W}P_{Y|W},Q_{XY|W}\right)\label{eq:-11-2}\\
 & =\psi_{s_{k_{i}}}\left(P_{W}P_{X|W}P_{Y|W},Q_{XY|W}^{\left(k_{i}\right)}\right).
\end{align}
Hence by the definition of $\psi_{s}\left(\cdot\right)$ in \eqref{eq:-35},
and the assumption $H(\widetilde{Q}_{XY|W}|P_{W})<H(P_{X|W}P_{Y|W}|P_{W})$,
we have that $\gamma_{k_{i}}$ diverges to $-\infty$ as $i\to\infty$.
However, in the RHS of \eqref{eq:-11-2}, by choosing $Q_{XY|W}$
as the specific distribution $P_{X|W}P_{Y|W}$, we know that $\gamma_{k_{i}}\ge I(XY;W)=C_{\mathsf{Wyner}}(X;Y).$
Hence the limit of $\gamma_{k_{i}}$ cannot be $-\infty$, which implies
equality in \eqref{eq:-10} holds, i.e., 
\begin{equation}
H(\widetilde{Q}_{XY|W}|P_{W})=H(P_{X|W}P_{Y|W}|P_{W}).\label{eq:-5}
\end{equation}
By \eqref{eq:-4} we know that \eqref{eq:-5} holds if and only if
$\widetilde{Q}_{XY|W}=P_{X|W}P_{Y|W}$. Hence 
\begin{align}
 & \Gamma_{1}^{\mathrm{UB}}(\pi_{XY})\nonumber \\
 & \leq\limsup_{i\to\infty}\sum_{w,x,y}P(w)Q^{\left(k_{i}\right)}(x,y|w)\log\frac{P(x|w)P(y|w)}{\pi\left(x,y\right)}\nonumber \\
 & \qquad+\frac{1}{s_{k_{i}}}\left(H(Q_{XY|W}^{\left(k_{i}\right)}|P_{W})-H(XY|W)\right)\label{eq:-3}\\
 & \leq\limsup_{i\to\infty}\sum_{w,x,y}P(w)Q^{\left(k_{i}\right)}(x,y|w)\log\frac{P(x|w)P(y|w)}{\pi\left(x,y\right)}\label{eq:-46}\\
 & =\sum_{w,x,y}P(w)\widetilde{Q}_{XY|W}(x,y|w)\log\frac{P(x|w)P(y|w)}{\pi\left(x,y\right)}\\
 & =\sum_{w,x,y}P(w)P(x|w)P(y|w)\log\frac{P(x|w)P(y|w)}{\pi\left(x,y\right)}\\
 & =C_{\mathsf{Wyner}}(X;Y),\label{eq:-7}
\end{align}
where \eqref{eq:-3} follows by the definition of $\Gamma_{1}^{\mathrm{UB}}(\pi_{XY})$
(below \eqref{eq:-35-4-1}) and \eqref{eq:-46} follows by \eqref{eq:-4}.

Combining \eqref{eq:-6} and \eqref{eq:-7}, we obtain that $\Gamma_{1}^{\mathrm{UB}}(\pi_{XY})=C_{\mathsf{Wyner}}(X;Y)$.

4) Proof of ``if'': If $\pi_{XY}$ satisfies the condition $(*)$,
then by \cite[Lemma 1]{yu2018on}, we have that $\Gamma_{\infty}^{\mathrm{UB}}(\pi_{XY})=C_{\mathrm{Wyner}}(\pi_{XY})$.
On the other hand, by Statement 2), $\Gamma_{1+s}^{\mathrm{UB}}(\pi_{XY})$
is non-decreasing in $s\in(0,\infty)$ and by Statement 3), $\Gamma_{1}^{\mathrm{UB}}(\pi_{XY})=C_{\mathrm{Wyner}}(\pi_{XY})$.
Hence $\Gamma_{1+s}^{\mathrm{UB}}(\pi_{XY})=C_{\mathrm{Wyner}}(\pi_{XY})$
for all $s\in(0,\infty]$.

Proof of ``only if'': This can be proven by a perturbation method,
which is similar to the proof of \cite[Lemma 1]{yu2018on}. Hence
we omit the proof. 

\bibliographystyle{unsrt}
\bibliography{ref}

\end{document}